\documentclass[preprint,aps ,nofootinbib]{revtex4}
\usepackage{graphicx}
\usepackage{amsmath}
\usepackage{amsfonts}
\usepackage{amssymb}
\usepackage{color}%
\usepackage{dcolumn}
\usepackage{slashed}

\providecommand{\U}[1]{\protect\rule{.1in}{.1in}}

\newcommand{\f}{\begin{equation}}
\newcommand{\ff}{\end{equation}}
\newcommand{\fa}{\begin{eqnarray}}
\newcommand{\ffa}{\end{eqnarray}}

\begin{document}

\title{Dynamic gap from holographic fermions in charged dilaton black branes}
\author{Jian-Pin Wu $^{1}$}
\email{jianpinwu@mail.bnu.edu.cn}
\author{Hua-Bi Zeng $^{2}$}
\email{zenghbi@gmail.com}
\affiliation{$^1$Department of Physics, Beijing Normal University, Beijing 100875, China\\$^2$Department of Physics, Nanjing University, Nanjing 210093, China}

\begin{abstract}
We report a dynamical
formation of a (Mott) gap from
holographic fermions.
By coupling a fermion field with dipole action to
the charged dilaton black branes
with a Lifshitz like IR geometry and $AdS_{4}$ boundary,
we find that when the dipole interaction is large enough, spectral weight
is transferred between bands, and beyond a critical dipole interaction,
a gap emerges in the fermion density
of states. The value of the gap becomes larger as the strength of
the interaction keeps increasing.

\end{abstract}

\maketitle

\section {Introduction}

As a new method to study strongly coupled
quantum field theory, AdS/CFT correspondence \cite{Maldacena1997,Gubser1998,Witten1998}
has been used to study the strongly coupled phenomena in quantum many-body
systems. One of the most successful progress is that by coupling
a free Fermion field to the \textbf{$Reissner-Nordstr\ddot{o}m$} (RN) black hole, a class of non-Fermi liquid
at zero temperature was found holographically\cite{HongLiuNon-Fermi,Sung-Sik Lee,HongLiuAdS2}.
The properties of this non-Fermi liquid in the presence of a magnetic field
have been studied in\cite{Magnetic1,Magnetic2,Magnetic3}.
The extension to finite temperature was investigated in\cite{FiniteT}. In \cite{BTZ,GBJPWu}, free fermions in AdS BTZ black hole and Gauss-Bonnet black hole have also been studied respectively. In addition, the fermions on Lifshitz Background have been explored in Refs.\cite{LifshitzFermions1,LifshitzFermions2}.
We can refer to the lecture \cite{HongLiuLecture} for an excellent latest review on this subjects.

In \cite{coupling1,coupling2}, the authors studied the holographic fermion system with
RN black hole background when a dipole interaction is added,
it is found that the strength of the interaction $p$
is similar to $U/t$ in the fermions Hubbard model\footnote{In Refs.\cite{coupling1,coupling2}, the authors construct the Pauli couplings in a bottom-up setup. The inspiration for using Pauli couplings in holographic
AdS/CMT came from the previous work\cite{upbottom1,upbottom2}, in which they derived the type of fermionic interactions that appear in the consistent supergravity truncations.}.
When $p$ is large enough, a gap opens just like in a
Mott insulator, in which a large repulsive
interaction $U$ will change the structure of the bands. Then the
fermion energy is in the middle of the gap, which makes an insulator\footnote{Another important paper on the effect dipole coupling is Ref.\cite{coupling3}, which mainly focus on the smaller value of dipole coupling. They find that the non-Fermi liquids behavior is robust under this deformation.}.

However, the RN black hole background has nonzero ground state entropy density, which
seems to be inconsistent with our intuition that a system of degenerate fermions has a unique
ground state. Therefore, a systematic exploration of the system that has zero extremal
entropy will be important and valuable. Such models have been proposed in Refs.\cite{ZeroEntropy1,ZeroEntropy2,ZeroEntropy3}.
Furthermore, in Ref.\cite{FermionsDilatonWu}, the author investigated the free fermionic response in the background proposed by Gubser and Rocha \cite{ZeroEntropy1}.
They find that the dispersion relation is linear, just like a Fermi liquid. It is very different from that found in RN black hole \cite{HongLiuNon-Fermi}.

Among these models of zero ground state entropy density, a very important and valuable model is that proposed by Goldstein et al. \cite{ZeroEntropy2}. Here the near horizon geometry is Lifshitz-like. Therefore, in this paper, we will study the characteristics of the fermionic response with dipole interaction in this background\footnote{The free fermionic response in the present background and
the fermionic response with dipole coupling in the background proposed by
Gubser and Rocha, will be addressed in two companion papers.}. The interesting results we find is that, similar to the fermions in the RN black hole background, the dipole interaction will open
a gap for the strongly coupled fermions system when the value of the interaction
$p$ is large enough.

The organization of this paper is as follows.
In section II, we first review the charged dilaton black branes with
zero entropy at zero temperature.
In section III, we derive the equations of motion
for the fermion field with a dipole action.
The main results of the emergence of gap is in section IV.
Finally, the discussion and conclusions are presented in section V.

\section{An extremal charged dilatonic black brane solutions}

\subsection{Einstein-Maxwell-dilaton model}

For completeness let us also review the dilaton gravity system \cite{ZeroEntropy2}.
Our discussion follows \cite{ZeroEntropy2} closely and some details are omitted.
The system we consider is as follows\footnote{We can also refer to Refs.\cite{dilatonextention1,dilatonextention2}, in which the holographic models of charged dilatonic black branes (without charged fluid) at finite temperature
and several classes of coupling functions are discussed in some details.}:
\begin{equation}
S=\frac{1}{2\kappa^2}\int
d^4x\sqrt{-g}[R-2(\partial\phi)^2-e^{2\alpha\phi}F^{ab}F_{ab}+\frac{6}{L^2}],
\end{equation}
where $R$ is Ricci scalar, $\phi$ is the dilaton field, $F_{ab}=\partial_{a}A_{b}-\partial_{b}A_{a}$ is
the field strength, and $L$ is the AdS radius.
In this paper we will focus on an
electrically charged black brane solution. The metric and gauge fields can be
written as,
\begin{equation}\label{DilatonMetric}
ds^2=-a^2(r)dt^2+\frac{dr^2}{a^2(r)}+b^2(r)[(dx)^2+(dy)^2],
\end{equation}
\begin{equation}\label{DilatonGauge}
e^{2\alpha\phi(r)}F=\frac{Q}{b^2(r)}dt\wedge dr.
\end{equation}
Such a gauge field satisfies the gauge field equation of motion automatically.
With the above ansatz, the remaining equations of motion can be expressed as follow:
\begin{equation}
(a^2b^2\phi')'=-\alpha e^{-2\alpha\phi}\frac{Q^2}{b^2},\label{KG}
\end{equation}
\begin{equation}
a^2b'^2+\frac{1}{2}(a^2)'(b^2)'=\phi'^2a^2b^2-e^{-2\alpha\phi}\frac{Q^2}{b^2}+\frac{3b^2}{L^2}.\label{CO}
\end{equation}
\begin{equation}
(a^2b^2)^{\prime\prime}=\frac{12b^2}{L^2}.\label{cx}
\end{equation}
\begin{equation}
\frac{b^{\prime\prime}}{b}=-\phi'^2.\label{ct}
\end{equation}
where the prime denote the differentiation with respect to the
coordinate $r$. In the subsequent subsections, we will construct such solutions that
the near-horizon geometry has a Lifshitz-like symmetry but at infinity the geometry asymptotes to AdS.
For convenience, we shall set $L=1$ and $\kappa=1$ in the following.

\subsection{Scaling solution near the horizon}

In order to construct the scaling solution near the horizon $r_h$, we consider the following ansatz:
\begin{equation}\label{ScalingSNH}
a=C_1r_{\ast}^\gamma,b=C_2r_{\ast}^\beta,\phi=-K\ln r_{\ast}+C_3,
\end{equation}
where $C_1$, $C_2$, $C_3$, $\gamma$, $\beta$, and $K$ are all
constants and $C_1$ and $C_2$ are set to be positive in the following, without loss of generality.
Note that for convenience, one has introduced the variable $r_{\ast}=r-r_h$ in the above ansatz.
With this ansatz and the equations of motion, we have
\begin{equation}
\gamma=1,\beta=\frac{(\frac{\alpha}{2})^2}{1+(\frac{\alpha}{2})^2},K=\frac{\frac{\alpha}{2}}{1+(\frac{\alpha}{2})^2},C_1^2=\frac{6}{(\beta+1)(2\beta+1)},
Q^2e^{-2\alpha C_3}=\frac{(2\beta+1)KC_1^2C_2^4}{\alpha}.
\end{equation}

Such a solution has a Lifshitz-like symmetry in the
metric, with a dynamical critical exponent $z=\frac{1}{\beta}$,
although such a symmetry is broken by the logarithmic dependence of
the dilaton on $r_{\ast}$. Also we note that in the above solution,
the metric component $g_{tt}$ has a double zero at the
horizon where $g_{xx}$ also vanishes and therefore corresponds to an
extremal brane with vanishing ground state entropy density.
Finally, we also point out that for $\alpha=0$, such a solution corresponds to an $AdS_{2}\times R^{2}$ geometry, which is also the near horizon geometry of the extremal RN black hole. The fermionic response in RN black hole has been explore carefully in Refs.\cite{HongLiuNon-Fermi,HongLiuAdS2}. Therefore we shall assume $\alpha>0$ and focus on the Lifshitz like near horizon geometry in the following.

\subsection{The solution with asymptotes to AdS at infinity}

To obtain a solution which asymptotes to AdS at infinity, we can add a perturbation to the scaling solution
which is irrelevant in the IR scaling region but relevant in the UV region.
In order to set the constant $C_{2}$ to unity and $C_{3}$ to vanishing,
we carry out a coordinate transformation under which,
\begin{equation}
r=\lambda\tilde{r},t=\frac{\tilde{t}}{\lambda},
x^i=\frac{{\tilde{x}}^i}{\bar{\lambda}}
\end{equation}
with $\lambda=e^{\frac{C_3}{K}}$ and
$\bar{\lambda}=\sqrt{C_2\lambda^\beta}$. Under such rescaling, we have
\begin{equation}
a=C_1r_{\ast},b=r_{\ast}^\beta,\phi=-K\ln r_{\ast},
\end{equation}
and in terms of $\alpha$, the charge can be expressed as
\begin{equation}
Q^2=\frac{6}{\alpha^2+2}.
\end{equation}

Subsequently, we add the perturbation to the Eqs. (\ref{ScalingSNH}).
The resulting functions satisfied the equations of motion at the leading order are
\begin{equation}\label{NHEOML}
a=C_1r_{\ast}(1+d_1r_{\ast}^\nu),b=r_{\ast}^\beta(1+d_2r_{\ast}^\nu),\phi=-K\ln r_{\ast}+d_3r_{\ast}^\nu,
\end{equation}
where $C_1$, $\beta$, and $K$ keep unchanged. The perturbation is characterized by the
exponent $\nu$ and the three constants $d_{1}$, $d_{2}$ and $d_{3}$. Also, $d_{2}$ and $d_{3}$ are determined by $d_{1}$ and they have the following relations: $d_3=\frac{2\beta+\nu-1}{2K}d_2$,
$d_1=[\frac{2(1+\beta)(1+2\beta)}{(2\beta+2+\nu)(2\beta+1+\nu)}-1]d_2$. In addition, the perturbation should die out at small $r_{\ast}$. Therefore, we must require $\nu> 0$, which gives rise to a unique allowed value for this exponent: $\nu=\frac{1}{2}[-2\beta-1+\sqrt{(2\beta+1)(10\beta+9)}]$. Thus there are two parameters $\alpha$ and $d_1$ for this perturbated solution. For simplicity, we will set $\alpha=1$.

\begin{figure}
\center{
\includegraphics[scale=0.9]{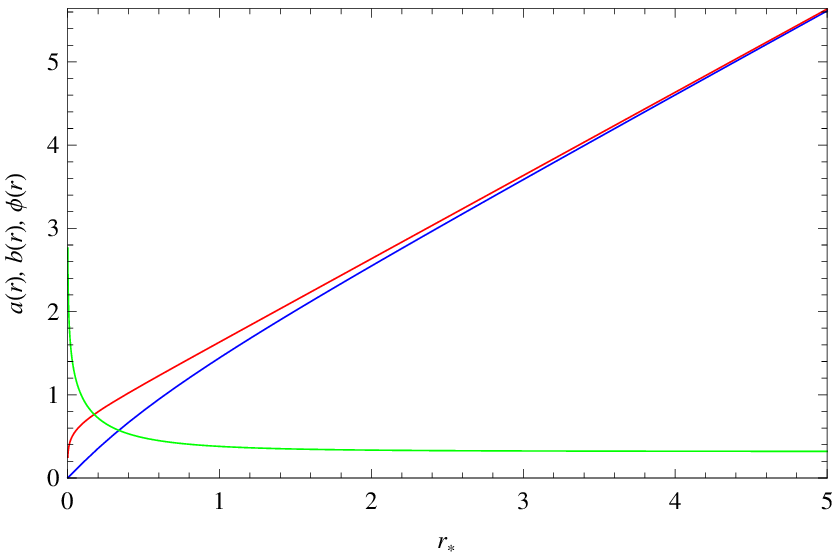}\hspace{0.4cm}
\includegraphics[scale=0.9]{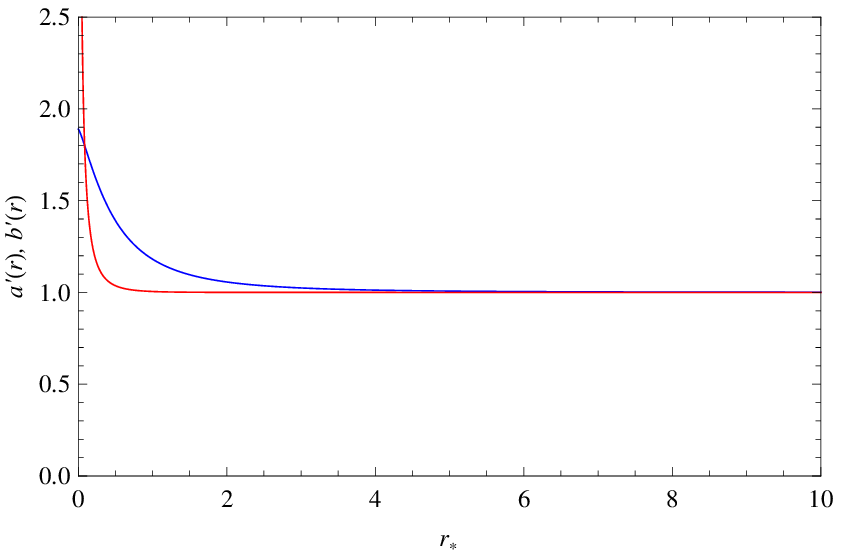}\\ \hspace{0.4cm}
\caption{\label{Background}Left Plot: Background solution in the case of $\alpha=1$ and $d_1=-0.514219$. Blue line denote $a(r)$, red line denote $b(r)$ and green line denote $\phi(r)$. Right plot: The plots for $a'(r)$ (blue line) and $b'(r)$ (red line). It indicates that $a'(r)$ and $b'(r)$ approach $1$ when $r$ approaches $\infty$.}}
\end{figure}

The initial data for numerical integration is taken from the above perturbated solution near the horizon.
As depicted in Fig. \ref{Background},
for the perturbation with $d_1=-0.514219$,
$b'(r)\rightarrow 1$ as $r\rightarrow \infty$, that is to say,
the solution near the boundary takes the standard asymptotics of AdS for the perturbation with $d_1=-0.514219$
\footnote{For any other negative value of $d_1$, $b$ does not take the standard form of
asymptotic AdS and we need a coordinate transformation to achieve our goal.
When $d_1>0$, the numerical solution is singular.}.
In the next section, we shall investigate the holographic fermions with dipole term
in such a background.

\section{Holographic fermion with bulk dipole coupling}

In order to know the effect of magnetic dipole coupling on the structure of
the spectral functions of the dual fermionic operator,
we consider the following bulk fermion action\footnote{Since the gauge field $A_{t}$
can be expressed as $A_{t}=\frac{a^{2}b^{2}\phi'}{Q\alpha}$
(It can easily be obtained by Eqs. (\ref{DilatonGauge}) and (\ref{KG})),
the bulk fermion action (\ref{actionspinor}),
only through the combination of ($qQ$,$pQ$), depends on ($q$,$p$). In this paper, for convenience, we only focus on the case of $q=4$ and $p\geq 0$. For more general parameter space ($q$,$p$), we will discusses them in the near future. When we have set $\alpha=1$, $Q=\pm \sqrt{2}$. Without loss of generality, we shall
take $Q=\sqrt{2}$ in the following. In this case, the chemical potential can be obtained as $\mu\simeq-0.843$ numerically. If we take $Q=-\sqrt{2}$, the same results will be recovered by tuning the parameter ($q$,$p$).
}
\begin{eqnarray}
\label{actionspinor}
S_{D}=i\int d^{4}x \sqrt{-g}\overline{\zeta}\left(\Gamma^{a}\mathcal{D}_{a} - m - ip\slashed{F}\right)\zeta,
\end{eqnarray}
where $\Gamma^{a}$ is related to the usual
flat space gamma matrix by a factor of the vielbein,
$\Gamma^{a}=(e_{\mu})^{a}\Gamma^{\mu}$, $\mathcal{D}_{a}=\partial_{a}+\frac{1}{4}(\omega_{\mu\nu})_{a}\Gamma^{\mu\nu}-iqA_{a}$
is the covariant derivative with $(\omega_{\mu\nu})_{a}$ the spin connection 1-forms
and $\slashed{F}=\frac{1}{4}\Gamma^{\mu\nu}(e_\mu)^a(e_\nu)^bF_{ab}$.
Before taking the particular background (Eqs. (\ref{DilatonMetric}) and (\ref{DilatonGauge})),
we will derive the flow equation as Refs.\cite{HongLiuNon-Fermi,GBJPWu,FermionsDilatonWu}
in a more general static background, i.e.,
\begin{eqnarray}
\label{GMetric}
ds^2=-g_{tt}(r)dt^2+g_{rr}(r)dr^2+g_{xx}(r)dx^2+g_{yy}(r)dy^2,~~~A_{a}=A_{t}(r)(dt)_{a}.
\end{eqnarray}
Firstly, the Dirac equation can be derived from the action $S_{D}$
\begin{eqnarray}
\label{DiracEquation1}
\Gamma^{a}\mathcal{D}_{a}\zeta-m\zeta -ip \slashed{F} \zeta=0.
\end{eqnarray}

Making a transformation
$\zeta=(-g g^{rr})^{-\frac{1}{4}}\mathcal{F}$ to remove the spin
connection and expanding $\mathcal{F}$ as $\mathcal{F}=F e^{-i\omega t +ik_{i}x^{i}}$
in Fourier space, the Dirac equation (\ref{DiracEquation1}) turns out to be
\begin{eqnarray}
\label{DiracEinFourier}
(\sqrt{g^{rr}}\Gamma^{r}\partial_{r}- m
- \frac{i p}{2} \sqrt{g^{rr}g^{tt}} \Gamma^{rt} \partial_{r}A_{t})F
-i(\omega+q A_{t})\sqrt{g^{tt}}\Gamma^{t}F
+i k \sqrt{g^{xx}}\Gamma^{x}F
=0,
\end{eqnarray}
where due to rotational symmetry in the spatial directions,
we set $k_{x}=k$ and $k_{i}= 0,~i\neq x$ without losing generality.
Notice that Eq. (\ref{DiracEinFourier}) only depends on three Gamma matrices $\Gamma^{r},\Gamma^{t},\Gamma^{x}$.
So it is convenient to split the spinors $F$ into $F=(F_{1},F_{2})^{T}$ and
choose the following basis for our gamma matrices as in \cite{HongLiuAdS2}:
\begin{eqnarray}
\label{GammaMatrices}
 && \Gamma^{r} = \left( \begin{array}{cc}
-\sigma^3  & 0  \\
0 & -\sigma^3
\end{array} \right), \;\;
 \Gamma^{t} = \left( \begin{array}{cc}
 i \sigma^1  & 0  \\
0 & i \sigma^1
\end{array} \right),  \;\;
\Gamma^{x} = \left( \begin{array}{cc}
-\sigma^2  & 0  \\
0 & \sigma^2
\end{array} \right),
\qquad \ldots
\end{eqnarray}

So, we have a new version of the Dirac equation as
\begin{eqnarray} \label{DiracEF}
(\sqrt{g^{rr}}\partial_{r}+m\sigma^3)\otimes \left( \begin{matrix} F_{1} \cr  F_{2} \end{matrix}\right)
=\left[\sqrt{g^{tt}}(\omega+qA_{t})i\sigma^2
\mp  k \sqrt{g^{xx}}\sigma^1
-p \sqrt{g^{tt}g^{rr}}\partial_{r}A_{t}\sigma^1\right]
\otimes \left( \begin{matrix} F_{1} \cr  F_{2} \end{matrix}\right).
~.
\end{eqnarray}

Furthermore, according to eigenvalues of $\Gamma^{r}$,
we make such a decomposition $F_{\pm}=\frac{1}{2}(1\pm \Gamma^{r})F$. Then
\begin{eqnarray} \label{gammarDecompose}
F_{+}=\left( \begin{matrix} \mathcal{B}_{1} \cr  \mathcal{B}_{2} \end{matrix}\right),~~~~
F_{-}=\left( \begin{matrix} \mathcal{A}_{1} \cr  \mathcal{A}_{2} \end{matrix}\right),~~~~
with~~~~F_{\alpha} \equiv \left( \begin{matrix} \mathcal{A}_{\alpha} \cr  \mathcal{B}_{\alpha} \end{matrix}\right).
\end{eqnarray}

Under such decomposition, the Dirac equation (\ref{DiracEF}) can be rewritten as
\begin{eqnarray} \label{DiracEAB1}
(\sqrt{g^{rr}}\partial_{r}\pm m)\left( \begin{matrix} \mathcal{A}_{1} \cr  \mathcal{B}_{1} \end{matrix}\right)
=\pm(\omega+qA_{t})\sqrt{g^{tt}}\left( \begin{matrix} \mathcal{B}_{1} \cr  \mathcal{A}_{1} \end{matrix}\right)
-(k \sqrt{g^{xx}}+p \sqrt{g^{tt}g^{rr}}\partial_{r}A_{t})
\left( \begin{matrix} \mathcal{B}_{1} \cr  \mathcal{A}_{1} \end{matrix}\right)
~,
\end{eqnarray}
\begin{eqnarray} \label{DiracEAB2}
(\sqrt{g^{rr}}\partial_{r}\pm m)\left( \begin{matrix} \mathcal{A}_{2} \cr  \mathcal{B}_{2} \end{matrix}\right)
=\pm(\omega+qA_{t})\sqrt{g^{tt}}\left( \begin{matrix} \mathcal{B}_{2} \cr  \mathcal{A}_{2} \end{matrix}\right)
+(k \sqrt{g^{xx}}-p \sqrt{g^{tt}g^{rr}}\partial_{r}A_{t}) \left( \begin{matrix} \mathcal{B}_{2} \cr  \mathcal{A}_{2} \end{matrix}\right)
~.
\end{eqnarray}

Introducing the ratios $\xi_{\alpha}\equiv \frac{\mathcal{A}_{\alpha}}{\mathcal{B}_{\alpha}},\alpha=1,2$,
one can package the Dirac equation (\ref{DiracEAB1}) and (\ref{DiracEAB2})
into the evolution equation of $\xi_{\alpha}$,
\begin{eqnarray} \label{DiracEF1}
(\sqrt{g^{rr}}\partial_{r}
+2m)\xi_{\alpha}
=\left[ v_{-} + (-1)^{\alpha} k \sqrt{g^{xx}}  \right]
+ \left[ v_{+} - (-1)^{\alpha} k \sqrt{g^{xx}}  \right]\xi_{\alpha}^{2}
~,
\end{eqnarray}
where $v_{\pm}=\sqrt{g^{tt}}(\omega+qA_{t})\pm p\sqrt{g^{tt}g^{rr}} \partial_{r}A_{t}$.
The above flow equation will be more convenient to impose the boundary conditions at the horizon
and read off the boundary Green functions.

Now, we take the particular background (Eqs. (\ref{DilatonMetric}) and (\ref{DilatonGauge})).
Because near the boundary, the geometry is an $AdS_{4}$,
the solution of the Dirac equation (\ref{DiracEF}) can be expressed as
\begin{eqnarray} \label{BoundaryBehaviour}
F_{\alpha} \buildrel{r \to \infty}\over {\approx} a_{\alpha}r^{m}\left( \begin{matrix} 0 \cr  1 \end{matrix}\right)
+b_{\alpha}r^{-m}\left( \begin{matrix} 1 \cr  0 \end{matrix}\right),
\qquad
\alpha = 1,2~.
\end{eqnarray}

If $b_{\alpha}\left( \begin{matrix} 1 \cr  0 \end{matrix}\right)$
and $a_{\alpha}\left( \begin{matrix} 0 \cr  1 \end{matrix}\right)$ are related by
$b_{\alpha}\left( \begin{matrix} 1 \cr  0 \end{matrix}\right)
=\mathcal{S}a_{\alpha}\left( \begin{matrix} 0 \cr  1 \end{matrix}\right)$,
then the boundary Green's functions $G$ is given by $G=-i \mathcal{S}\gamma^{0}$ \cite{HongLiuSpinor}.
Therefore
\begin{eqnarray} \label{GreenFBoundary}
G (\omega,k)= \lim_{r\rightarrow \infty} r^{2m}
\left( \begin{array}{cc}
\xi_{1}   & 0  \\
0  & \xi_{2} \end{array} \right)  \ ,
\end{eqnarray}

Near the horizon, we take the scaling solution near the horizon (\ref{ScalingSNH}).
We find that the requirement that the solutions of Eqs. (\ref{DiracEAB1}) and (\ref{DiracEAB2})
near the horizon be in-falling implies
\begin{eqnarray} \label{GatTip}
\xi_{\alpha}\buildrel{r \to r_{h}}\over =i,~~~~for~~\omega\neq 0.
\end{eqnarray}

\section{Emergence of the gap}

\begin{figure}
\center{
\includegraphics[scale=0.58]{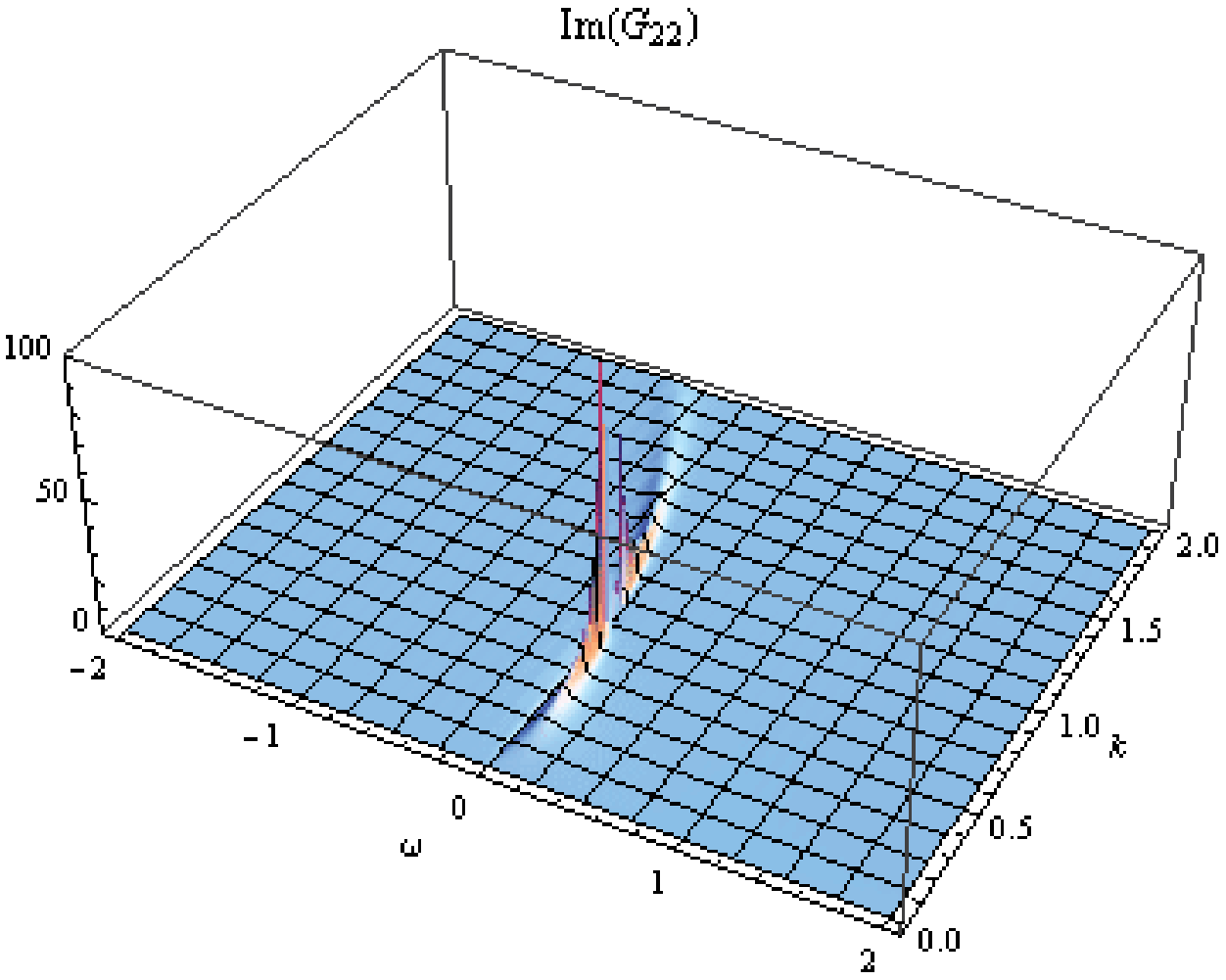}\hspace{0.4cm}
\includegraphics[scale=0.58]{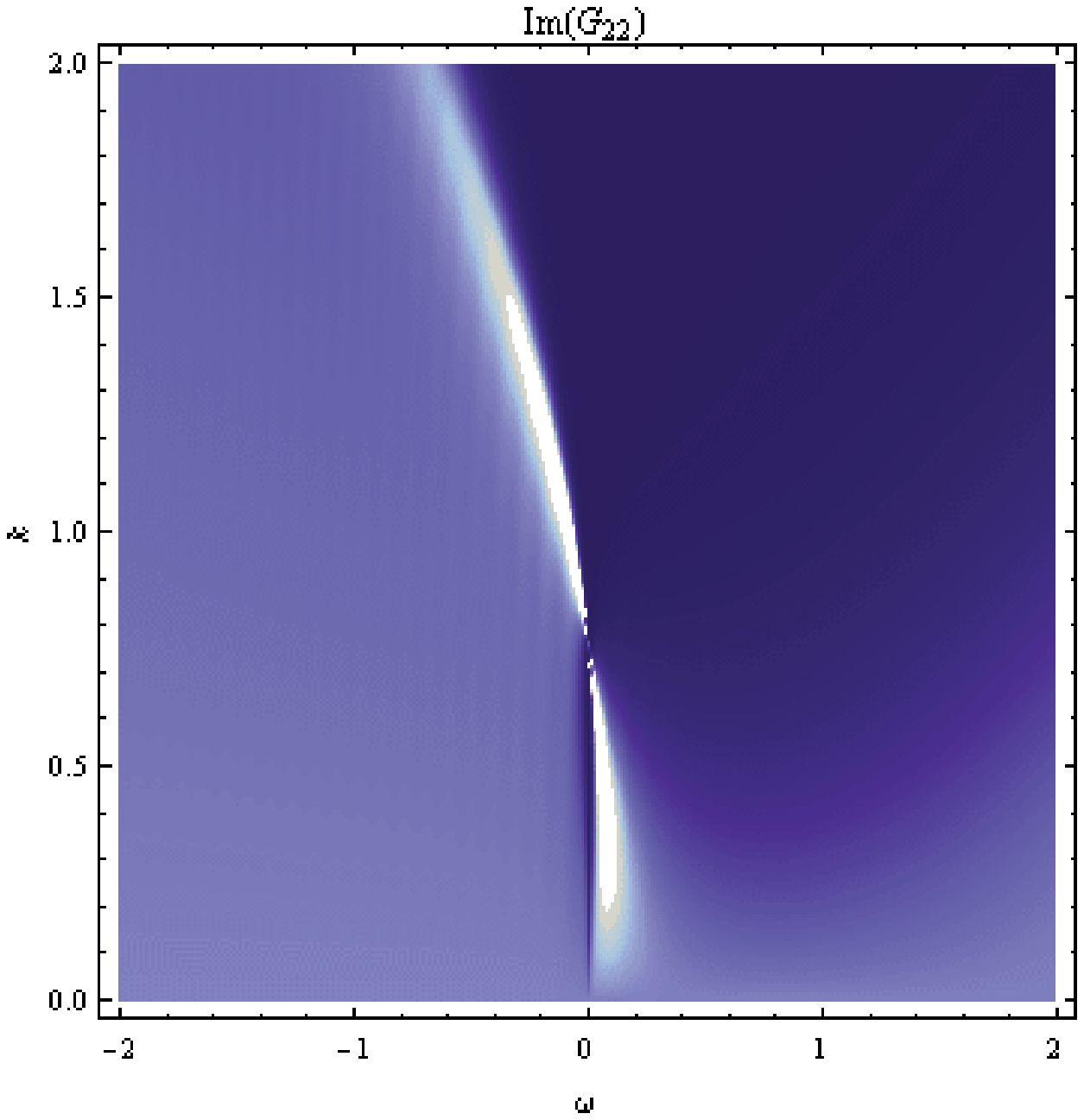}\\ \hspace{0.4cm}}
\center{
\includegraphics[scale=0.58]{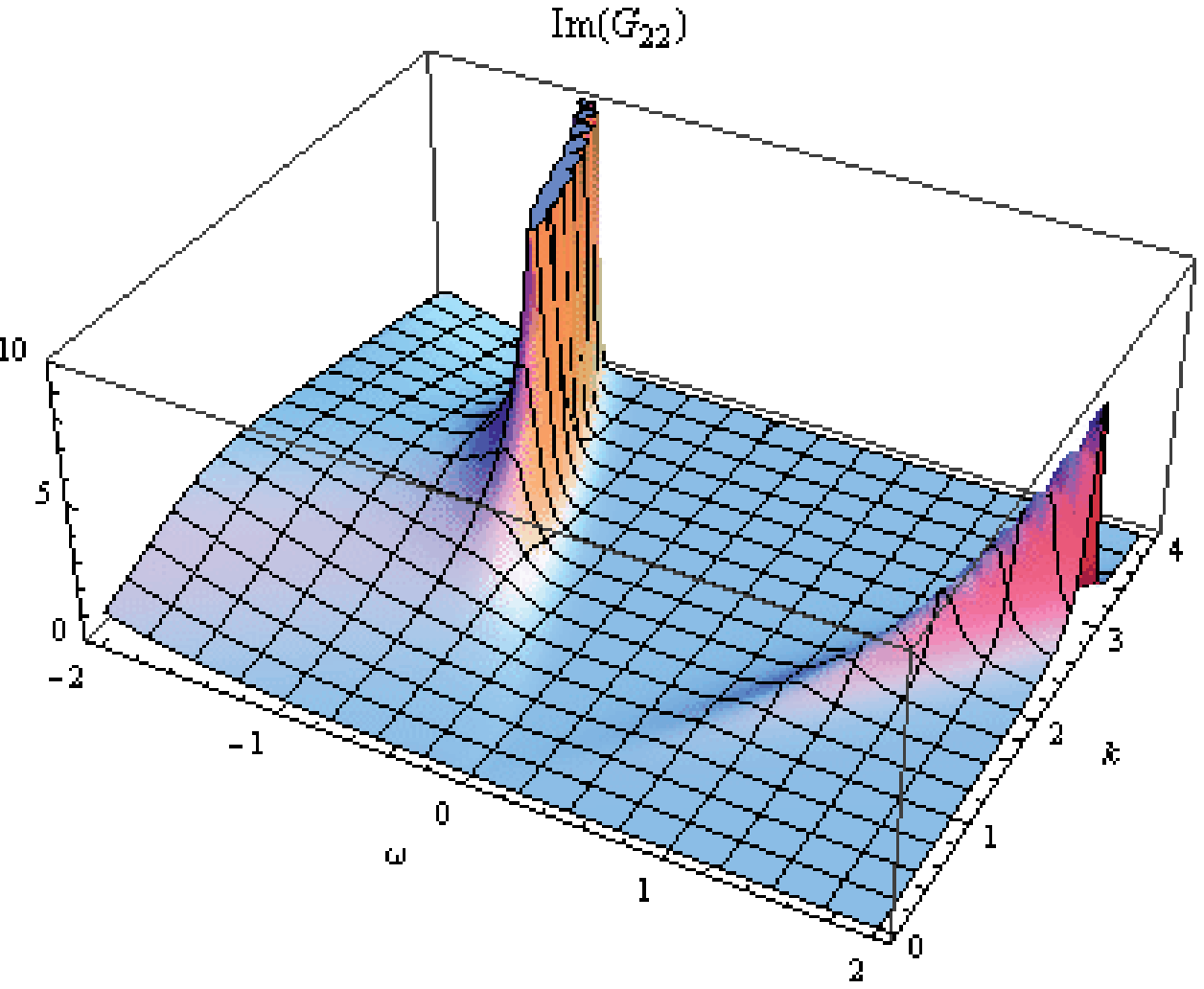}\hspace{0.4cm}
\includegraphics[scale=0.58]{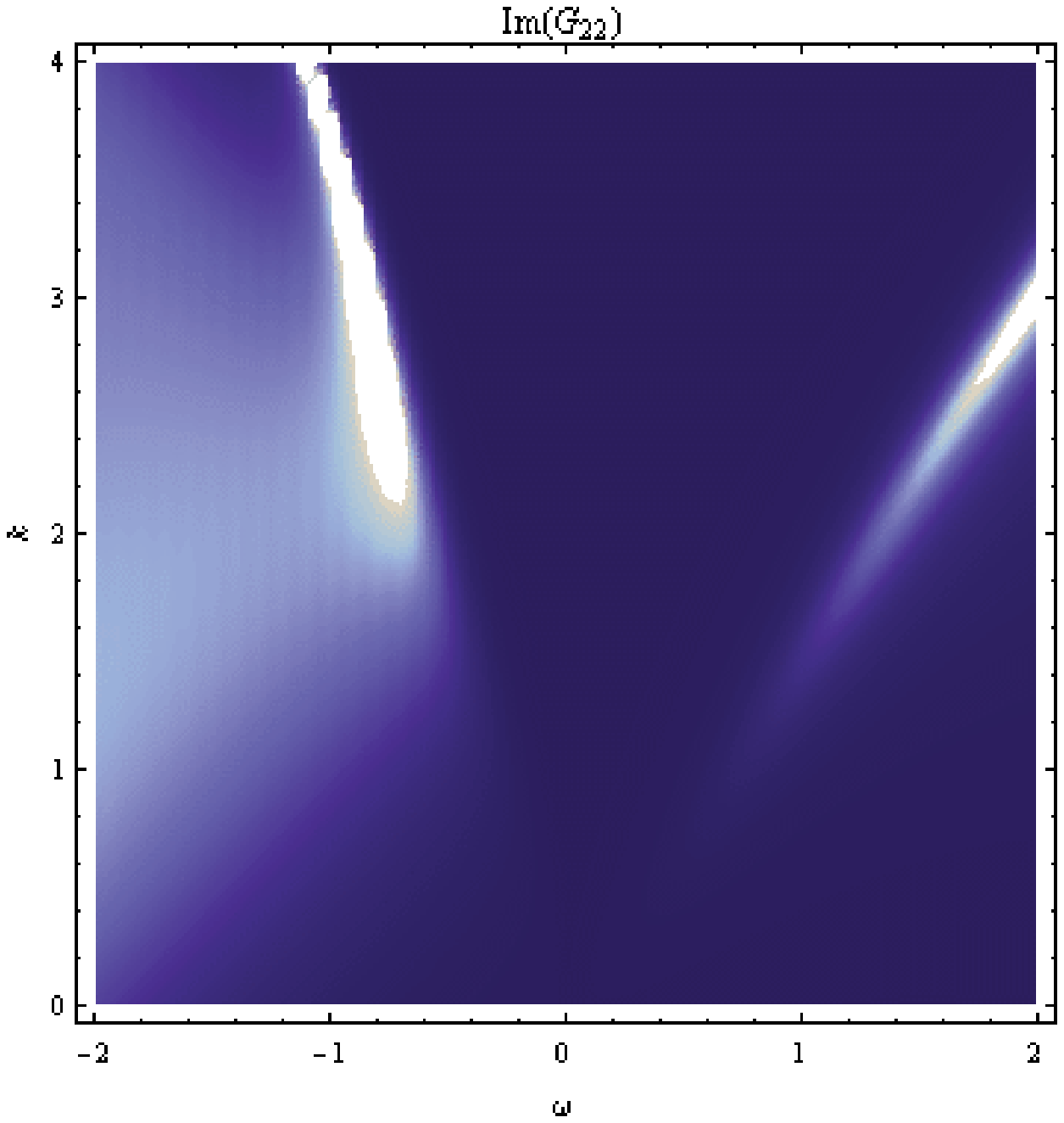}\\ \hspace{0.4cm}
\caption{\label{3DDp0p2}The 3d and density plots of $ImG_{22}(\omega,k)$ for $q=4$ and $m=0$
(the plots above for $p=0$ and the plots below for $p=2$.).
In the plots above, a quasiparticle-like peak occurs at $k\simeq 0.75$ and $\omega=0$.
However, in the plots below, a gap emerges around $\omega=0$.
}}
\end{figure}

In this section, to study the spectral function, we will numerically solve
the flow equation (\ref{DiracEF1}) with the initial condition (\ref{GatTip}).
Up to normalization, the spectral function is given by $A(\omega,k)=Im[TrG_{R}(\omega,k)]$.
However, due to the relation $G_{11}(\omega,k)=G_{22}(\omega,-k)$,
$G_{11}(\omega,k)$ can be recovered from $G_{22}(\omega,k)$.
Therefore, in the following 3D plots and density plots, to make numerical calculation easier,
we will only show $G_{22}(\omega,k)$.

\subsection{The Fermi momentum $k_{F}$}

Before discussing the spectral function for the larger $p$,
we would like to give a simple discussion on the Fermi momentum $k_{F}$ for $p=0$.
From the plots above in FIG.\ref{3DDp0p2}, one can see that for $p=0$,
a sharp quasiparticle-like peak occurs near $k\simeq 0.75$ and $\omega=0$,
indicating a Fermi surface. When $q$ is increased, the Fermi momentum $k_{F}$ increases (Table I).
As $q$ is further increased, several Fermi surfaces emerge.
For example, for $q=6$, the another Fermi surface begin to occur at $k\simeq 0.168$.
In fact, the case that with the increase of $q$, several Fermi surfaces produce also occurs in RN-AdS background.
It seems that the gravity duals with larger charge $q$ posses more branches of Fermi surfaces.
Conversely, with the decrease of $q$, the Fermi sea gradually disappears.
\begin{widetext}
\begin{table}[ht]
\begin{center}

\begin{tabular}{|c|c|c|c|c|c|c|c|}
         \hline
~~$q$~~ &~~$3$~~&~~$3.5$~~&~~$4$~~&~~$4.5$~~&~~$5$~~&~~$5.5$~~&~~$6$~~
          \\
        \hline
~~$k_{F}$~~ &~~$0.2481$~~&~~$0.4874$~~&~~$0.7565$~~&~~$1.0457$~~&~~$1.3494$~~&~~$1.6643$~~&~~$1.9878$~~
          \\
        \hline
\end{tabular}
\caption{\label{kF} The Fermi momentum $k_{F}$ for different charge $q$.}

\end{center}
\end{table}
\end{widetext}

In addition, we would also like to point out that for fixed $q=-1$ and $\alpha=10^{-6}$, $k_{F}\simeq 0.917$, which recover the results of Ref.\cite{HongLiuNon-Fermi}\footnote{Since we take the negative chemical potential here, $q=-1$ corresponds to $q=1$ in Ref.\cite{HongLiuNon-Fermi}. In addition, for $\alpha=0$, the numerical solution is singular. Therefore, in the numerical calculation, we take $\alpha=10^{-6}\simeq 0$ instead of $\alpha=0$.}. For more discussions on the Fermi momentum $k_{F}$ and the low energy behaviors in this case, we will explore it in the another companion paper.

\subsection{Emergence of the gap}

\begin{figure}
\center{
\includegraphics[scale=0.58]{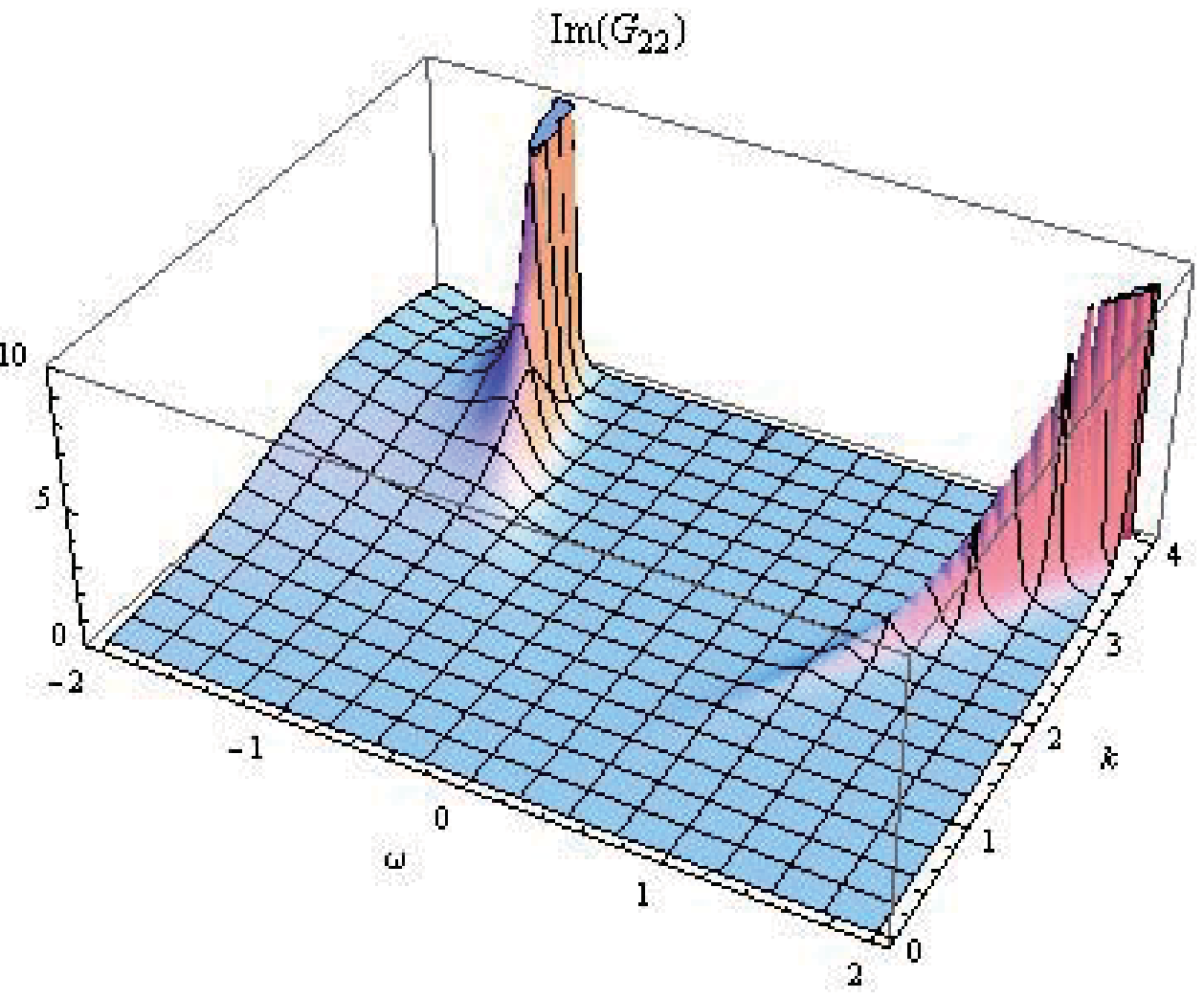}\hspace{0.4cm}
\includegraphics[scale=0.58]{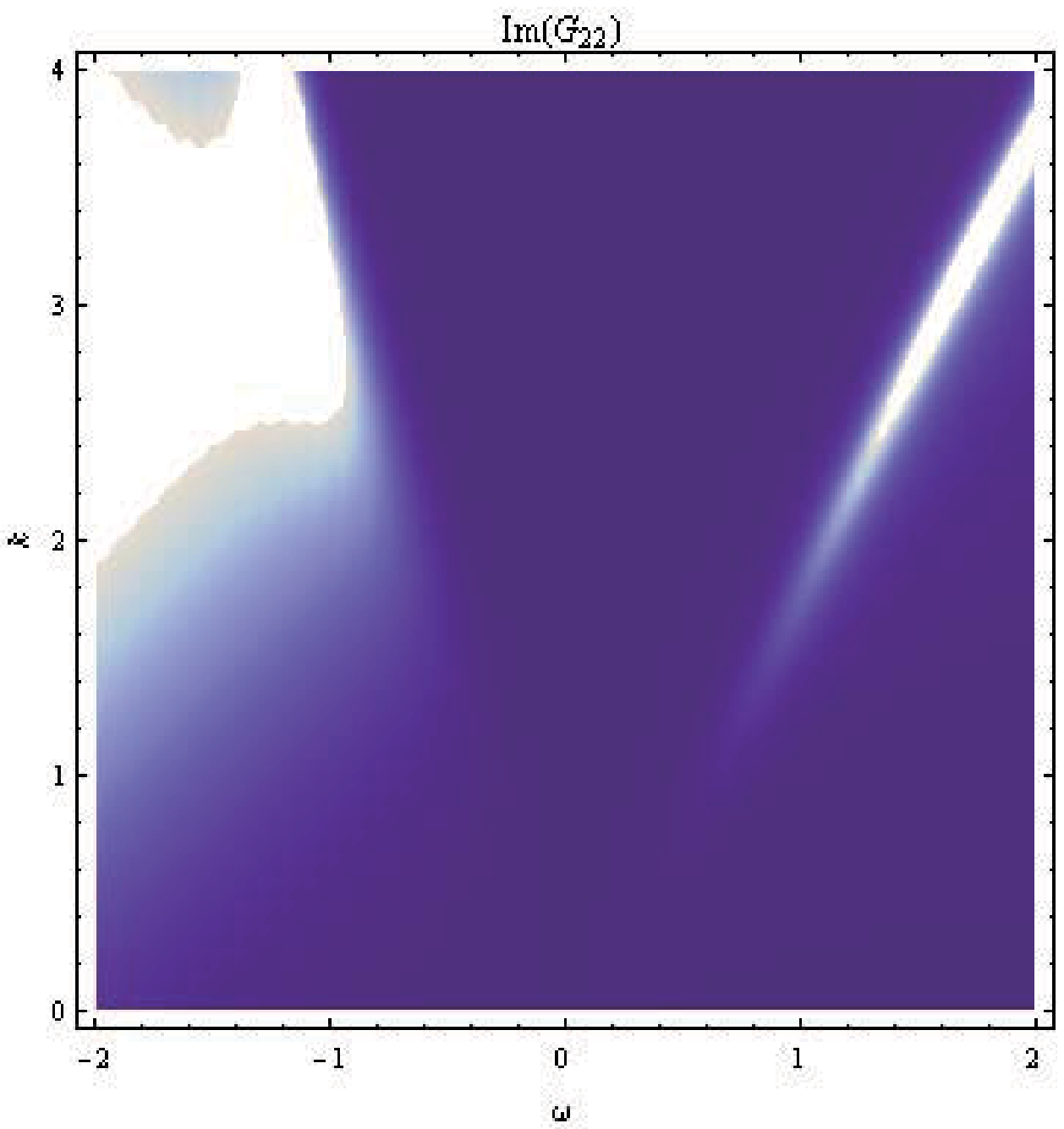}\\ \hspace{0.4cm}}
\center{
\includegraphics[scale=0.58]{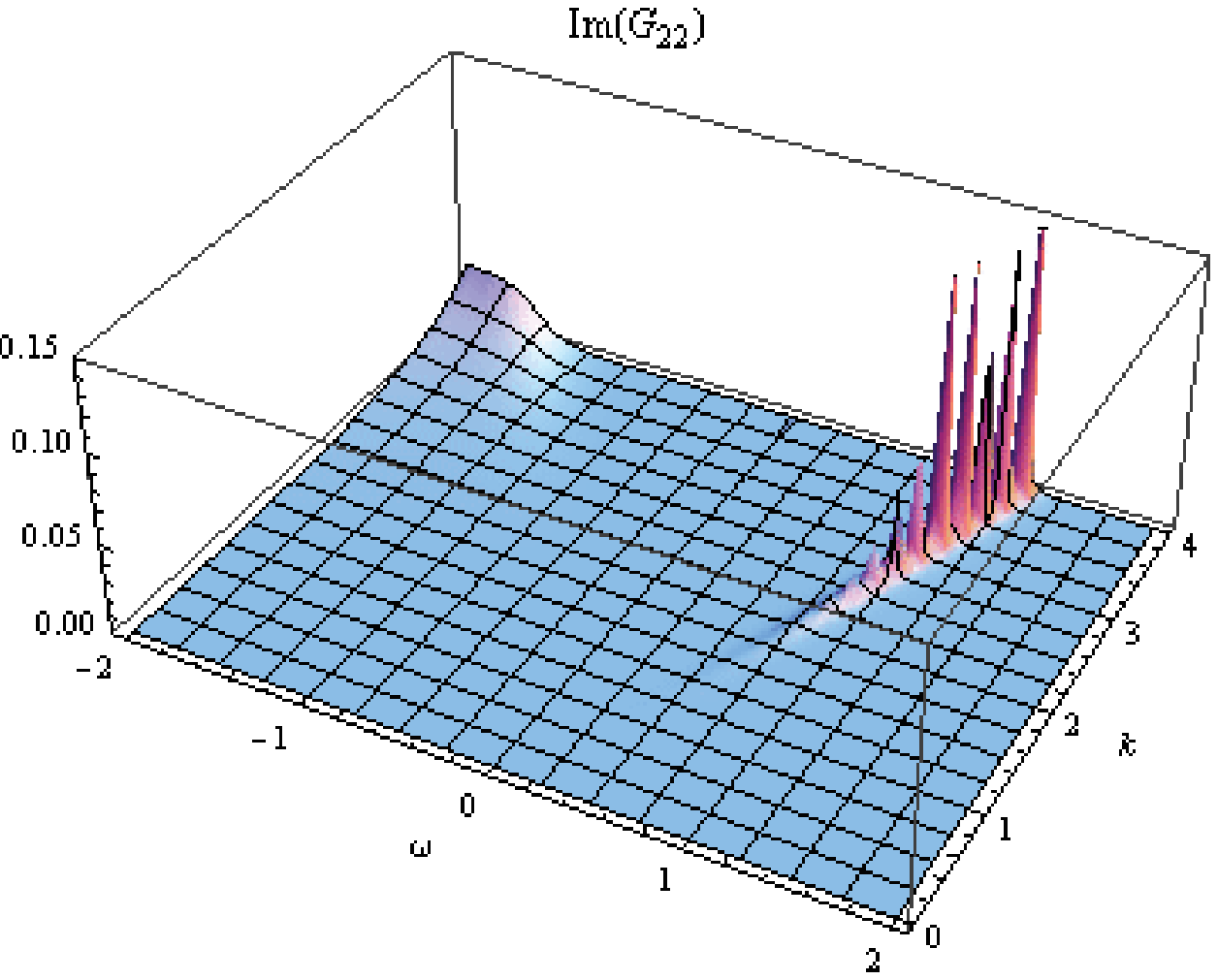}\hspace{0.4cm}
\includegraphics[scale=0.58]{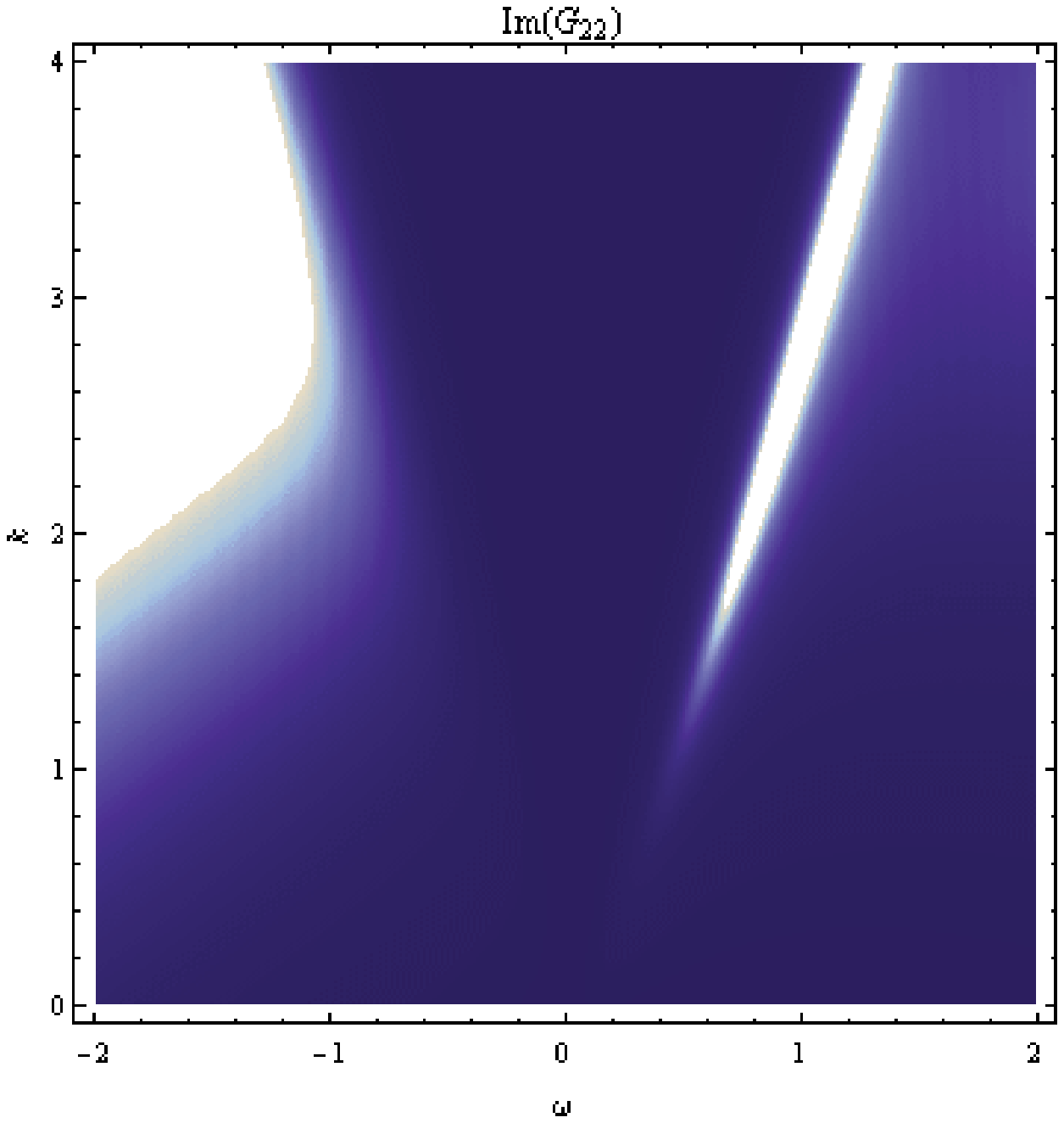}\\ \hspace{0.4cm}
\caption{\label{3DDp25p45}The 3d and density plots of $ImG_{22}(\omega,k)$ for $q=4$ and $m=0$
(the plots above for $p=2.5$ and the plots below for $p=4.5$.).
}}
\end{figure}

When we turn on $p$ to the value of $2$, a gap emerges and
there are two bands located at the positive frequency
and negative frequency regions, respectively (the plots below in FIG.\ref{3DDp0p2}).
Evidently, the strength of the lower band is bigger than the upper band.
We also show the 3D plot and density plot of
the Green's function $ImG_{22}$ for $p=2.5$ and $p=4.5$ in FIG.\ref{3DDp25p45}.
We can see that with the increase of $p$, the width of the gap becomes larger
and the band in the negative frequency region switch gradually to the positive frequency region.
As $p$ increases further, the lower band disperses and the upper band becomes stronger and sharper.

\begin{figure}
\center{
\includegraphics[scale=0.88]{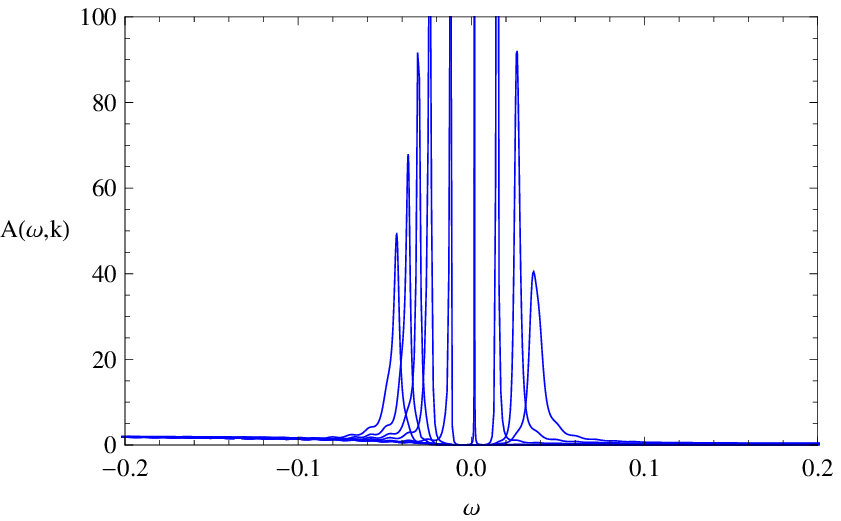}\hspace{0.2cm}
\includegraphics[scale=0.88]{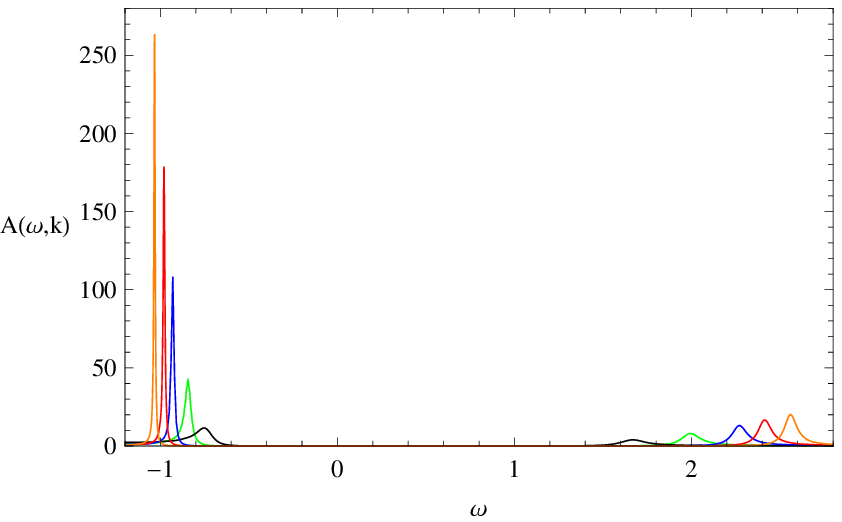}\\ \hspace{0.2cm}}
\center{
\includegraphics[scale=0.88]{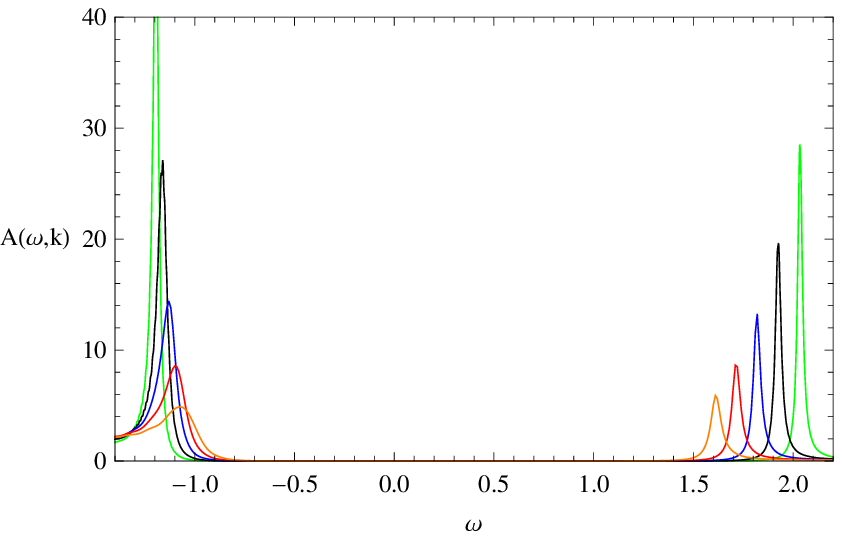}\hspace{0.2cm}
\includegraphics[scale=0.88]{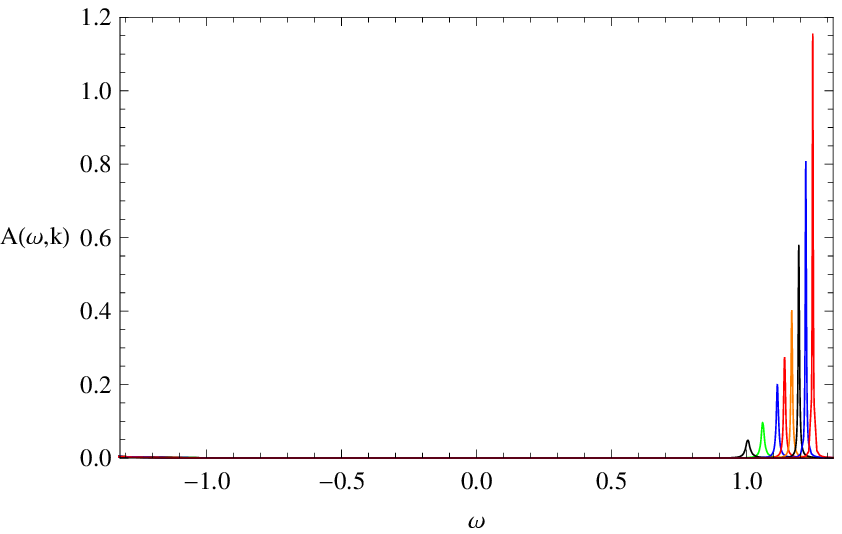}\\ \hspace{0.2cm}
\caption{\label{Ap}The plots of spectral function $A(\omega,k)$
as a function of $\omega$ for sample values of $k\in [0.6,3.8]$
for $p=0$ (left plot above), $p=2$ (right plot above),
$p=2.5$ (left plot below) and $p=4.5$ (right plot below).
For $p=0$, the sharp quasiparticle-like peak occurs at $\omega=0$
when we dial $k=k_{F}\simeq 0.75$. However, for $p=2$, $2.5$ and $4.5$,
the gap around $\omega=0$ persists for all $k$.}}
\end{figure}

\begin{figure}
\center{
\includegraphics[scale=0.88]{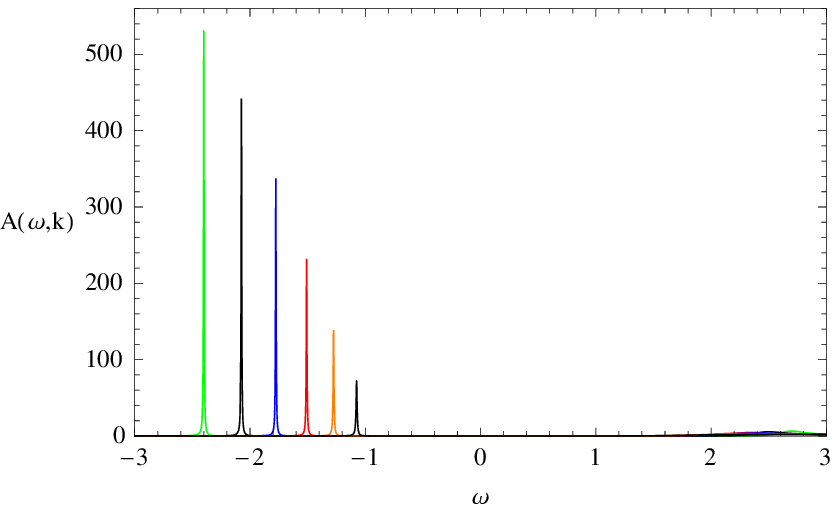}\hspace{0.2cm}
\includegraphics[scale=0.88]{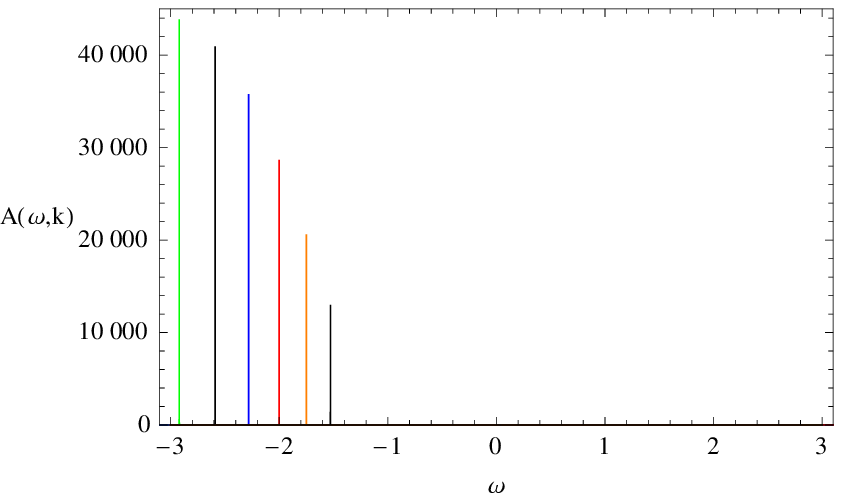}\\ \hspace{0.2cm}
\caption{\label{RNAp}For the background of RN black hole, the plots of spectral function $A(\omega,k)$
as a function of $\omega$ for sample values of $k\in [1,4.5]$
for $p=6$ (left plot), $p=8$ (right plot).
}}
\end{figure}

In order to explore further the characteristics of the spectral function,
We show the spectral function $A(\omega,k)$ for $p=0$, $2$, $2.5$ and $4.5$
for sample values of $k$. From the left plot above in FIG.\ref{Ap},
we find that some peaks distribute at both positive and negative frequency regions.
When the peak approaches $\omega=0$, its height goes to infinity, and its width goes to zero,
indicating that there is a Fermi surface at $k=k_{F}$.
When we amplify $p$, the strength of the quasiparticle-like peak at $\omega=0$ degrades,
and vanishes at a critical value $p_{crit}$. For $p>p_{crit}$,
the dipole interaction will open a gap as the case in RN black hole \cite{coupling1}.
Differ from the case of $p<p_{crit}$, the height of the peaks degrade
and the width becomes larger when $\omega=0$ is approached, and the peak vanishes around $\omega=0$.
We also note that the spectral density mainly appears
at negative frequency regions for $p=2$ (the right plot above in FIG.\ref{Ap}).
As $p$ increases, the spectral density begins to distribute the positive frequency
and negative frequency regions, respectively (the left plot below in FIG.\ref{Ap}).
When $p$ increases further, the spectral density switches gradually to
the positive frequency regions (the right plot below in FIG.\ref{Ap})\footnote{It seems that
for the case of the RN black hole, the spectral density always appears at negative frequency regions ($p>0$,FIG.\ref{RNAp}).}.

\begin{figure}
\center{
\includegraphics[scale=0.9]{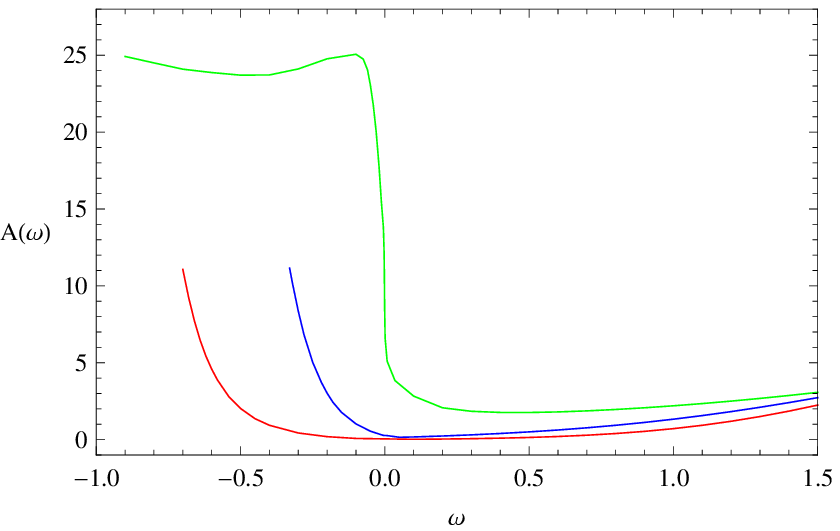}\hspace{0.4cm}
\includegraphics[scale=0.9]{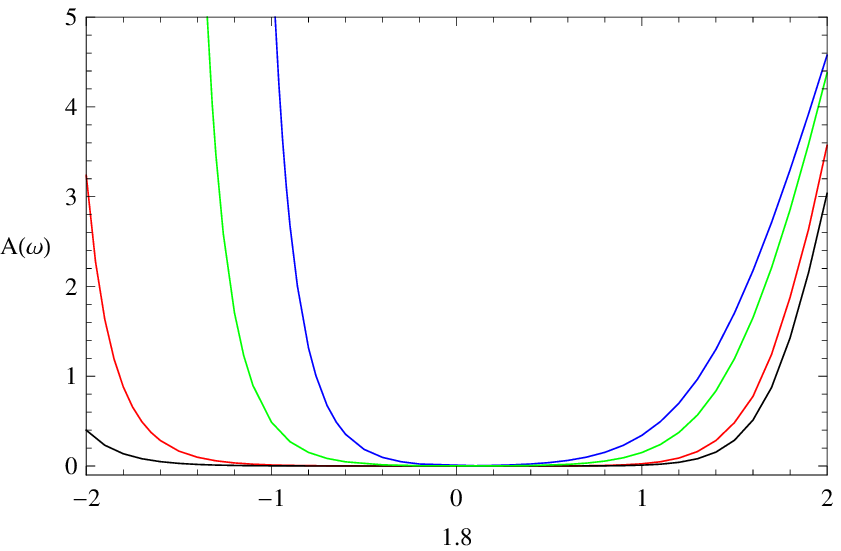}\\ \hspace{0.4cm}
\caption{\label{Density}Left plot: The density of states $A(\omega)$
for $p=1$ (green), $p=1.5$ (blue) and $p=2$ (red);
Right plot: The density of states $A(\omega)$
for $p=2.5$ (blue), $p=3$ (green), $p=4$ (red) and $p=4.5$ (black).
The onset of the gap is at $p\simeq 1.5$.}}
\end{figure}

The another important quantity of interest for us is the density of states $A(\omega)$,
which is defined as the integral of $A(\omega,k)$ over $k$.
It is the total spectral weight.
By careful numerical computations, we find that the onset of the gap is near $p=1.5$,
which is different from the onset value of $p=4$ in the case of RN background.
Consistent with the above observations (FIG.\ref{3DDp0p2}, FIG.\ref{3DDp25p45} and FIG.\ref{Ap}),
the total spectral weight is also mainly distribute at negative frequency regions
for small $p$ ($p>p_{crit}$, left plot in FIG.\ref{Density}),
and switches to the positive frequency regions as the increase of $p$.

Finally, we show the relation between the width of the gap $\Delta$ and $p$.
Evidently, the gap becomes wider as $p$ increases.

\begin{figure}
\center{
\includegraphics[scale=0.9]{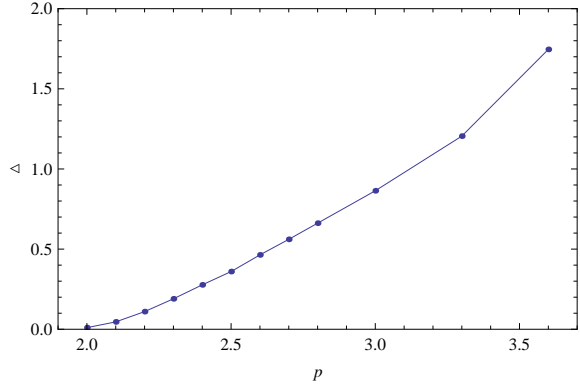}\\ \hspace{0.4cm}
\caption{\label{gapwidth}The width $\Delta$ of the gap as a function of $p$.}}
\end{figure}

\section{Conclusions and discussion}

In this paper, we have explored the fermionic response in the presence of a bulk dipole interaction term with
strength $p$ in the dilatonic black brane with a Lifshitz like IR geometry and $AdS_{4}$ boundary. When $p=0$, a sharp quasiparticle-like peak occurs near $k=0.75$ and $\omega=0$, indicating a Fermi surface. As $p$ increases, the peak become wide and its strength degrades gradually, and the spectral weight is transferred between bands. Furthermore, when $p$ goes beyond a critical dipole interaction $p_{crit}$, the Fermi sea disappears and a gap emerges as the case in RN black hole \cite{HongLiuNon-Fermi}. By studying the another important quantity: the density of states $A(\omega)$, we clearly find that the onset of the gap is near $p=1.5$ and the width of the gap becomes bigger as $p$ increases.

So far, we have found a (Mott) gap from holographic fermions in the background of the dilatonic black brane with a Lifshitz like IR geometry when a dipole interaction term are added as that found in RN black hole. However, for more general dilatonic black brane, which also have vanishing ground entropy density, for example, a class of systems proposed in Refs.\cite{ZeroEntropy1,dilatonicBB1,dilatonicBB2,dilatonicBB3}, is the emergence of gap robust? The fermionic response without dipole interaction in this class backgrounds has been investigated in Refs.\cite{FermionsDilatonWu,FermionsDilatonG}. Therefore, it is valuable to
test the robustness of the emergence of the gap in the backgrounds of more general dilatonic black
branes. In addition, it is also interesting to extend our investigations in this paper to the case of the non-relativistic fermionic fixed point. We will address them in the near future.

\begin{acknowledgments}

We especially thank Professor Robert Leigh for his very valuable correspondence.
J.P. Wu would also like to thank Hongbao Zhang and Wei-Jia Li for their collaboration in the related project.
We also thank the referee for many valuable comments and correcting some English language typos.
J.P. Wu is partly supported by NSFC(No.10975017) and the Fundamental Research Funds for the central Universities.
H.B. Zeng is supported by the Fundamental Research Funds for the Central Universities (Grant No.1107020117) and
the China Postdoctoral Science Foundation (Grant No. 20100481120).

\end{acknowledgments}

\end{document}